\documentclass[11pt,twoside]{article}

\usepackage{asp2006}
\usepackage{fancyhdr,graphicx,caption,rotating,multirow,lscape}
\DeclareGraphicsExtensions{.eps,.eps.gz,.ps,.jpg,.tiff}

\begin{document}

\def\eq#1{equation (\ref{#1})}
\def\Eq#1{Eq.~\ref{#1}}
\def\mjup{M_{Jup}}
\def\rjup{R_{Jup}}
\def\days{{\rm days}}
\def\be{\begin{equation}}
\def\ee{\end{equation}}
\def\sn{{\rm S/N}}
\def\nave{\langle N \rangle}

%%% Fill in title
\title{Statistics and Simulations of Transit Surveys for Extrasolar Planets}  

%%% Fill in author names, use initials and surname and affiliation
%%% One author
\author{B. Scott Gaudi}
\affil{Department of Astronomy, The Ohio State University, 140 W. 18th Ave., Columbus, OH 43210}

\begin{abstract}
The yields from transit surveys can be used to constrain the frequency
and statistical properties of extrasolar planets. Conversely, planet
frequencies can be used to estimate expected detection rates, which
are critical for the planning and execution of these surveys. Here I
review efforts to accomplish these two related goals, both of which
generally require realistic simulations. Early
attempts to predict planet yields generally resulted in overly
optimistic detection rates that have not been realized.  I point out
where these estimates likely went wrong, and emphasize the
strong biases and sensitivity to detection thresholds inherent in
transit surveys.  I argue that meaningful comparisons between observed
and predicted detection rates require proper calibration of these
biases and thresholds.  In the few cases where this has been done, the
observed rates agree with the results from radial velocity surveys for
similar stellar environments.  I then go on to describe recent,
detailed calculations which should provide more accurate rates, which
can be critically compared to observed yields. Finally, I discuss
expectations for future all-sky synoptic surveys, which
may have the sensitivity to detect hundreds or thousands of close-in
transiting planets.  Realizing the enormous potential of these surveys
will require novel methods of coping with the overwhelming number of
astrophysical false positives that will accompany the planet detections.
\end{abstract}

\section{Introduction} 

There are two basic reasons for conducting transit surveys for
extrasolar planets. The first, most obvious, and most familiar, it
simply to find transiting planets in order that one can perform the
host of follow-up studies that are possible with these systems.  Such
studies enable one to measure many otherwise unobservable physical
properties of the planets (see \citealt{charbonneau06} and references
therein).  As these studies are best suited to bright systems,
uncovering the transiting planets orbiting the brightest host stars is
the primary motivation of many wide-angle photometric surveys, as well
as several radial velocity surveys.

The second reason to conduct transit surveys is to constrain the
frequency of short-period planets as a function of mass, radius, and
period.  Since only $\sim 10\%$ of short-period planets
transit their host stars, and the transit duty cycle is only $\sim
5\%$, one might wonder whether it is wiser to
do this using a detection method which
shows a persistent signal over a larger range of inclinations, such as
precision radial velocities.  Of course this is being done, but currently
photometric surveys can probe a larger number of systems over a larger
range of distances from the Sun, and so can detect intrinsically
rarer systems, or constrain the properties of planets in environments
beyond the local solar neighborhood, such as open clusters, globular
clusters, the distant Galactic disk, or even the Galactic bulge.

Proper planning and execution of
transit surveys requires the ability to predict the number of planets
that will be detected, given a model for the frequency and
distribution of planets, the properties of the survey, and the
detection criteria.  Conversely, these predictions are required in
order to use observed detection rates from completed surveys to
infer the intrinsic frequency and distribution of planets.  Here I
review efforts to accomplish these two related goals,
both of which generally require realistic simulations 
of transit surveys.

\section{Predicting Planet Yields}\label{sec:plan}

One can broadly divide the methods of predicting planet yields into two
categories.  The first category, which I will call reverse (or {\it a
posteriori}) modeling, uses the known properties of an observed sample
of stars around which one is searching for planets to calibrate the
survey efficiency.  The second category, which
I will call forward (or {\it ab initio}) modeling, uses assumed
distributions of the stellar and planetary properties to statistically
predict the ensemble properties of the target stars and the detection
efficiency of the survey as a whole.  These two methods have different
advantages and drawbacks, and their relative usefulness depends on
the context in which they are applied.

In the reverse approach, one attempts to model, as accurately as
possible, the total detection probability of each individual star in
the survey sample.  The expected number of detections is then just the
sum of the detection probabilities over all the stars in the survey.
For example, given a known mass $M_k$ and radius $R_k$ for each star
$k$, one can determine the individual transit probabilities $P_{{\rm
tr},k}$.  These stellar properties can be combined with the properties
of the survey observations (cadence, photometric uncertainties,
correlations) to determine the detection probability for each star $P_{{\rm
det},k}$.  Then, assuming a distribution $df/drdP$ of planets as
function of period $P$ and planet radius $r$, the (differential) number of expected
planet detections, $\nave$, is given by,
\be
\frac{d\nave}{dr dP}
= \frac{df(r,P)}{dr dP}  \sum_k P_{{\rm tr},k}(M_k,R_k,P)
P_{{\rm det},k}(M_k,R_k,r,P).
\label{eqn:nback}
\ee

The advantage of the reverse approach is that it is more
accurate and accounts for Poisson fluctuations in the individual
stellar properties. The disadvantage is that it requires knowledge of
the properties of the individual stars.  This is relatively
straightforward to obtain for surveys toward stellar systems such as globular or
open clusters, but can be quite difficult to obtain for field surveys, for which
the physical properties of the individual stars are poorly constrained due to
their unknown distances and foreground extinction.  Furthermore, the
reverse approach is not generally applicable for predicting the
expected yields of future surveys.

In the forward approach, one dispenses with any hope of modeling the 
detection probabilities of the stars individually, but rather attempts
to construct statistical distributions of the stellar properties, and 
use these to predict the ensemble detection probability of the stars
in the survey.  The average number of
planets that a transit survey should detect is the product of
the differential stellar density distribution along the line-of-sight 
($dn/dMdRdL$), the distribution of planets ($df/drdP$), the
transit probability, and the detection probability:
\be
\frac{d\nave}{dM dR dL dr dP d\ell d\Omega}
= \frac{d n(\ell,M)}{dM dR dL}
\frac{d f(r,P)}{dr dP}
P_{tr}(M,R,P) P_{\rm det}(M,R,L,r,P) \ell^2.
\label{eqn:dngen}
\ee
Here $L$ is the stellar luminosity, $l$ is the distance
along the line-of-sight, and thus $\ell^2d\ell d\Omega$ represents the volume
element. 

Of the ingredients that enter into \eq{eqn:dngen}, the detection
probability is one of the most critical and must be specified
with care. In particular, the detection probability must account for
{\it all} selection cuts that are employed in
the actual surveys, such as cuts on 
parameters output from transit-finding algorithms\footnote{For example,
the $\alpha$ and SDE parameters
in the BLS algorithm \citep{kovacs02}.}, the number of transits,
the source color and magnitude, and the radial velocity precision (as used
for confirmation).  When comparing the predictions of surveys to
results from completed experiments, the cuts {\it must} be applied
consistently in the data and the model, otherwise any inferences
are highly suspect.  This can be particularly difficult when trying to model
any `by-eye' cuts.

The forward approach is well-suited to the planning of future surveys,
and field surveys where the masses and radii of the individual stars
are not known.  The drawback of the forward approach is that it is
less exact, requires considerably more input assumptions, and is sensitive
to uncertainties in these assumptions.  One can
improve the accuracy of the forward approach by adopting a hybrid
method in which one imposes external observational constrains on the
model.  For example, one of the most important indicators of the
expected number of detections is simply the total number of stars
being surveyed.  Therefore, observed number counts of stars as a
function of magnitude and color can be used to constrain the parameters of
the forward model and so make the predicted yields more reliable.

\subsection{Simple Estimates Fail}

The previous discussion notwithstanding, 
at first glance transit surveys appear fairly straightforward, and
the requirements for detecting a transiting planet may seem reasonably
clear. Given that the transit probability  for
short-period planets is $P_T \sim 10\%$,
given that transits are expected to have a depth of $\delta \sim 1\%$, and given
that the frequency of short period planets is known to be $f\sim 1\%$,
then one might expect the number of detected events to be 
\be
\nave \sim f P_{\rm tr} N_{\le 1\%} \sim 10^{-3} N_{\le 1\%},\qquad {\rm (Naive\,\,  Estimate)},
\label{eqn:nsimp}
\ee
where $N_{\le 1\%}$ is the number of surveyed stars with photometric precision better than
$\sigma \sim 1\%$.  Early estimates of the yield of transit surveys
varied in complexity, but generally centered around this extreme
simplification of \eq{eqn:dngen}.  This estimate would imply that
one would need to monitor  only $\sim 10^3$ stars with $\sigma \la 1\%$ for 
a duration of $2P\sim 6~\days$ in order to detect a transiting planet.  In fact,
results from successful surveys imply that the number of stars that must be monitored is closer to 
$\sim 10^5$, depending on the survey.

Why does this simple estimate fail?  There are many reasons, but the
most important include the following: (1) A large fraction (often the
majority) of the stars in the field are either giants or upper
main-sequence stars that are too large to enable the detection of transits due to 
Jupiter-sized planets \citep{gould03,brown03}.  (2) 1\% photometry is
not a sufficient requirement for detecting a transit, one typically
needs to exceed some sort of signal-to-noise ratio $(\sn)$ requirement. The
\sn\ in turn depends on the depth of the transit, the photometric accuracy, and
the number of points taken during transit. Furthermore, 
uncertainties in ground-based photometry can be correlated on the
typical time scales of transits, thereby reducing the statistical
power of the data \citep{pont06}. (3) One generally requires several
transits for detection, which when combined with the small duty cycle of
the transits and losses in single-site observations, can result in a substantial suppression
of the number of planets detected. (4) Magnitude-limited radial velocity surveys
are more biased toward metal-rich (and so planet-rich) stars than $\sn$-limited transit surveys.
Therefore the frequency of planets is likely to be substantially smaller than 1\% for the typical
stars monitored in transit surveys \citep{gould06}.

\subsection{Selection Effects are Critical}

One key point that must be
addressed when simulating transit surveys is what it means to `detect' a transiting planet. 
Typically
transit surveys use a number of criterion to select transit-like
features from observed light curves, but most trigger on some variant
of a $\sn$ criterion\footnote{For example, the
$\alpha$ parameter in the BLS search algorithm
\citep{kovacs02} is often used to select the best transit candidates,
and is closely related to the $\sn$.}.  For uncorrelated noise, the
$\sn$ is approximately given by, \be \frac{S}{N} \simeq
N_{\rm tr}^{1/2}\frac{\delta}{\sigma},
\label{eqn:sngen}
\ee where $N_{tr}$ is the total number of points in transit.  Under
simple assumptions (Poisson-noise limited photometry of the source, no
extinction, random sampling) this can be written in terms of physical
parameters as \citep{gaudi05,gaudi05b}, 
\be \frac{S}{N} \propto
[R^{-3/2}M^{-1/6}L^{1/2}]\, [r^2 P^{-1/3}]\, \ell^{-1}.
\label{eqnsnprop}
\ee  Given a minimum $(\sn)_{\rm min}$, one can invert this equation
to determine the distance $\ell_{\rm max}$ out to which one can detect a
given planet orbiting a given star.

The expected number of planets detected around a uniform population of
stars is \citep{pepper03},
\be 
\nave \sim
\frac{\Omega}{3} n P_{\rm tr} \ell_{\rm max}^3,
\label{eqn:nave}
\ee
where $n$ is the volume density of stars, and $\Omega$ is the area
of the field-of-view.  At a limiting $\sn$, it is
then relatively straightforward to show that, 
\be
\nave \propto P^{-5/3}r^6 \left(\frac{S}{N}\right)_{\rm min}^{-3}.
\label{eqn:navescale}
\ee
Although trivial to derive, this expression has profound
implications for transit surveys.  For example, it implies that
$\sn$-limited transit surveys are $\sim (1/3)^{-5/3}\sim 6$ times more
sensitive to planets with $P\sim 1~{\rm day}$ than planets with $P\sim
3~\days$.  This effect alone largely explains why the first planets
uncovered by transit surveys had periods that were shorter than
any found in radial velocity surveys \citep{gaudi05}.  Also, the
extremely strong scaling with planet radius, $\propto r^6$, implies
that transit surveys will always detect the extreme, bloated planets
first; this bias must be properly considered when interpreting the
radius distribution of observed planets \citep{gaudi05b}.  Finally,
the number of detected planets is expected to be a strong function of the limiting
$\sn$, and thus survey teams must carefully specify their $\sn$ limit in
order to assess their expected yield.  This is not necessarily trivial
because, although they may initially use automated cuts with rigid thresholds to select
candidates, subsequent rejection of candidates by visual inspection
imposes a higher (and more difficult to model) $\sn$ threshold.

\section{Constraints on the Frequency of Planets}

A number of groups have used transit surveys to measure or constrain
the frequency of short-period giant planets based on the results of 
completed surveys.  The majority of these studies have
focused on cluster environments, where it is easier to determine the
properties of the target stars.  

The first such study was the now-famous HST survey of 47 Tuc
\citep{gilliland00}, which found no transiting planets and used this
null result, combined with an estimate of the survey efficiency, to
demonstrate that the frequency of short-period planets was more than
an order of magnitude smaller than in the local solar neighborhood.
This conclusion was confirmed and strengthened by \citet{weldrake05},
who further argued that metallicity was the likely cause for this
difference in the planet population.

A number of groups have searched for planets in open clusters, without
success.  Several of these groups have calibrated their detection
efficiencies and used these to place weak constraints on frequency of
short-period planets in these systems \citep{mochejska05,mochejska06,bramich06,
burke06}.

Few groups have attempted to use the results from field surveys to
constrain the frequency of short period planets.  As mentioned
previously, field surveys are complicated by the fact that the source stars
are located at a range of distances and suffer from a range of
extinctions, and therefore the relevant parameters of the target stars
(e.g. their radii and mass) are not known simply from their observed
fluxes and colors.  Because of this, forward modeling provides the best method of
estimating the efficiencies of field surveys.

%One figure
\begin{figure}[!ht]
\centering
\includegraphics[angle=0,width=7.5cm]{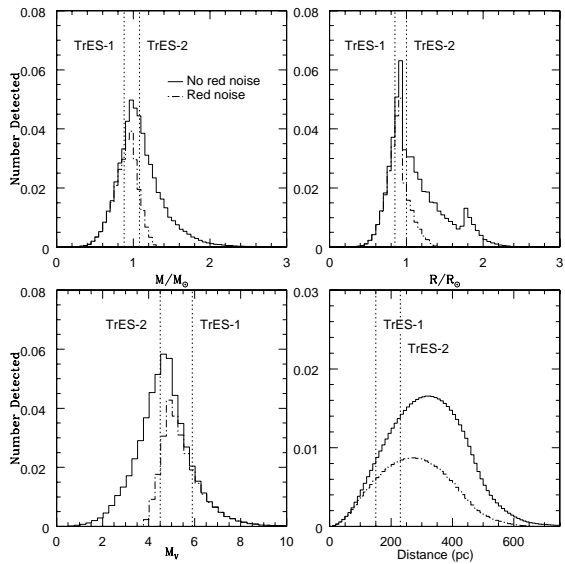}
\caption{
The distribution of predicted TrES detections in the Lyr1 field
as a function of the primary mass ({\bf Top Left}), 
radius ({\bf Top Right}), distance ({\bf Bottom Right}), and 
absolute $V$-magnitude ({\bf Bottom Left}) .  The solid lines are for uncorrelated uncertainties,
the dashed-dot line is for uncertainties correlated at the $0.3\%$
level.  The location of the stars TrES-1 and
-2 are also shown. TrES-2 lies within Lyr1, while TrES-1 is in the
nearby Lyr0 field. From \citet{beatty07}.\label{fig:tres}}
\end{figure}

In a comprehensive study, \citet{gould06} modeled the expected yield
of the first two campaigns of the OGLE transit survey
\citep{udalski02a,udalski02b,udalski02c,udalski03},
taking careful account of the survey selection effects, and using a
detailed model for the population of source stars.  They then used
the five planets detected in the OGLE survey to infer that the
fraction of stars with planets is $(1/710)\times 1^{+1.10}_{-0.54}$
for $P=1-3~\days$ and $(1/320)\times 1^{+1.37}_{-0.59}$ for $P=3-5~\days$,
consistent with the results from RV surveys.  They noted, however,
that magnitude-limited RV surveys are biased toward metal-rich (and so
planet-rich) stars, while transit surveys are not, therefore one would
generally expect to find a deficit of planets in transit surveys in
comparison to RV surveys.  \citet{gould06} also demonstrated
that the sensitivity of the OGLE surveys declined rapidly for $r\la \rjup$, indicating
that little can be said about the frequency of sub-Jovian
sized planets.

\section{Predictions for Ongoing and Future Surveys}\label{sec:sim}

Several authors have developed models of the expected yields of
transit surveys, with various levels of complexity.  \citet{horne03}
derived a simple analytic expression for the yields of pencil-beam transit
surveys and applied this to several ongoing projects, 
whereas \citet{pepper03} presented and applied the formalism
for estimating the yields of all-sky surveys.  \citet{pepper05}
developed a model to predict the number of detected planets in
surveys of stellar systems; this was significantly extended and improved
upon by \citet{aigrain06}. \citet{brown03} included, for the first time,
expectations for the rates of false alarms as well, and \citet{gillon05} used a
detailed model to estimate and compare the potential of several space-based and
ground-based surveys.

\begin{table}
\caption {Predicted TrES Yields ($R\leq13$ and $r=1\rjup$)}
\begin{center}
{\small
\begin{tabular}{c|ccc}
\tableline
\tableline
Period & $1-3~\days$ & $3-5~\days$ & Both\\  
\tableline
S/N$\geq10$ & 3.96 & 4.47 & 8.43\\
S/N$\geq15$ & 3.23 & 3.30 & 6.53\\
S/N$\geq20$ & 2.53 & 2.35 & 4.88\\
S/N$\geq25$ & 1.94 & 1.66 & 3.60\\
S/N$\geq30$ & 1.49 & 1.19 & 2.68\\
\tableline
\end{tabular}
}
\end{center}
\label{tab:tres}
\end{table}

\citet{beatty07} attempt to build upon and advance these previous
studies, accounting for as many real-world effects as possible, including
the variation of the stellar density along
the line-of-sight, a $\sn$ detection criterion, various
noise sources (source, sky, scintillation, saturation, uncorrelated
and correlated\footnote{See \citet{pont06}.} uncertainties), apparent
magnitude limits, the stellar mass function, the magnitude-scale height
relation, requirements on the minimum number of detected transits, and
arbitrary bandpasses.  

Figure \ref{fig:tres} shows predictions
for the number of transiting planets detected by the TrES
survey \citep{dunham04,alonso04,odonovan06} toward one of their target
fields in Lyra, assuming the frequencies of giant planets from
\citet{gould06}.  The field center, number of observations, and
photometric errors were taken from the TrES
website\footnote{http://www.astro.caltech.edu/$\sim$ftod/tres/sleuthObs.html}.  
For this field, $\nave=0.6$ detections are
expected for planets with $r=1.0\rjup$ for $R\le 13$ and $\sn>20$,
assuming uncorrelated noise.  Assuming noise correlated at the level
of $\sim 0.3\%$, the number of detections generically drops by a
factor of $\sim 2$.  Over the $\sim 10$ TrES fields which have been
exhausted for planets, a total of $\nave \sim 5$ detections are
expected in this model.  This compares reasonably well with the actual
yield of two detections. The remaining discrepancy could be because
correlated noise is important, or the effective $\sn$ is higher.
Table \ref{tab:tres} shows the number of detections as a function of
the limiting \sn.  For example, for $\sn>30$, the number of detections
drops to $\nave \sim 2.7$.  This further demonstrates the important
point that, in order to use the observed yield of TrES (or any other
transit survey) to infer the frequency of short period planets, it is
essential to accurately characterize the limiting detection
threshold.

One important ingredient that is missing in most of the previous
simulations is the ability to predict the rate of false positives,
including grazing eclipsing binaries, unrelated blends, and
hierarchical triples.  The rates for these astrophysical false
positives are expected to be even higher than the rate of bona fide
planet detections.  The one exception is the detailed model of
\citet{brown03}, however this model did not include some of the
important effects considered by others, such as density variations
due to Galactic structure or
a $\sn$ detection criterion.  A more sophisticated model
that incorporates all the relevant effects is needed.

%One figure
\begin{figure}[!ht]
\centering
\includegraphics[angle=0,width=7.5cm]{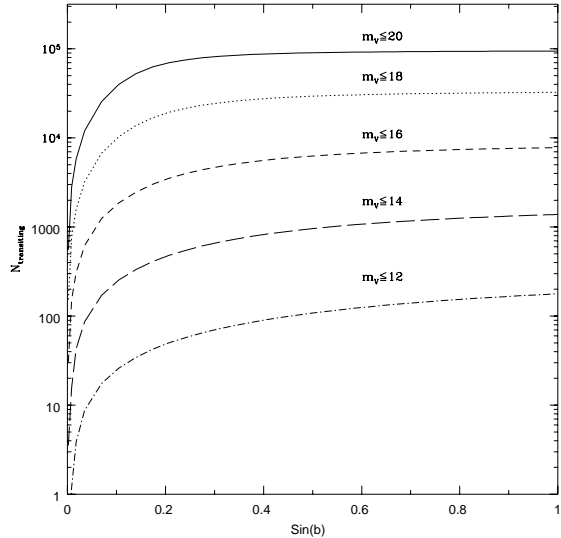}
\caption{
The cumulative number of transiting, Jupiter-sized, short-period ($P=1-5~{\rm days}$)
planets around Sun-like stars as a function of Galactic latitude,
for various limiting $V$ magnitudes. From \citet{beatty07}.\label{fig:allsky}}
\end{figure}

\subsection{The Potential of All-Sky Synoptic Surveys}

One difficulty with trying to understand the trends that are emerging
among the population of transiting planets is simply the small size of
the sample.  Ideally, one would like to be able to subdivide the
sample and ask, e.g., how the trends depend on the mass of the primary
star.  For such analyses, a sample size of a least hundred transiting
planets will be required.  This an order of magnitude larger than the
number of transiting planets known today.  Figure \ref{fig:allsky}
shows the cumulative number of transiting, Jupiter-sized, short-period
($P=1-5~\days$) planets orbiting solar type stars as a function of
Galactic latitude for several different limiting $V$ magnitudes, based
on the simulations of \citet{beatty07}.  There are $\nave \sim 200$
such planets over the whole sky down to $V=12$.  Unfortunately,
current wide-field surveys are unlikely to survey a sufficient
fraction of the sky to detect more than few dozen of these, given that
the most ambitious of these projects only monitor $\sim
10\%$ of the sky.

Increasing the number of known transiting planets by an order of
magnitude will likely require a fundamentally different approach, or
at least a significant upgrade to the current experiments.  Plans are
being made to this end, but it is interesting to ask what the
potential is to detect transiting planets in the large scale synoptic
surveys that are being currently being built or planned (and that are
{\it not} specifically designed to detect transiting planets). 

\begin{table}
\caption {Predicted Yields for Large Synoptic Surveys ($r=\rjup, P=3~\days$)}
\begin{center}
{\small
\begin{tabular}{c|cccc}
\tableline
\tableline
                & \multicolumn{2}{c}{Sun-like} &\multicolumn{2}{c}{ M-dwarfs}\\
                & $V_{lim}$ & $\nave$ & $V_{lim}$ & $\nave$ \\
\tableline      
LSST            & 18.5 &  7740  & 23.1  & 15530\\
SDSS-II         & 15.6 &  6.0   & 20.2  &  11.9\\
Pan-STARRS      & 15.0 &  19.2  & 19.6  &  36.5\\
Pan-STARRS Wide & 12.5 &  48.0  & 17.1  & 81.6\\
\tableline
\end{tabular}
}
\end{center}
\label{tab:allsky}
\end{table}

An accurate estimate of the yield of large synoptic surveys requires
careful simulations.  This can be difficult,
since in some cases the relevant parameters of the experiments have
not been finalized.  However, we can make a crude
estimate by combining the
predictions from Figure \ref{fig:allsky} with a rough estimate of the
limiting magnitude of the survey.  Assume a given setup with diameter
$D=D_0$ can achieve a (source-noise limited) precision of
$\sigma=\sigma_0$ on a star with $V=V_0$ with an exposure time of
$t_0$. Then, for uncorrelated noise, the limiting magnitude of stars
around which a survey can detect transiting planets with
$\sn>(\sn)_{\rm min}$ is,
\begin{equation}
V_{lim} = 5\log{\left[ \left( \frac{ \epsilon T}{t_0}
\frac{\Omega}{\Theta} \frac{R}{\pi a} \right)^{1/2} \frac{D}{D_0}
\frac{\delta}{\sigma_0} \left(\frac{S}{N}\right)_{\rm
min}^{-1}\right]} + V_0,
\end{equation}
where $T$ is the total duration of the experiment, $\Theta$ is the
total area surveyed, $\Omega$ is the field-of-view of the camera, and
$a$ is the semimajor axis of the planets, and $\epsilon$ is the total
survey efficiency (fraction of $T$ spent exposing).

Table \ref{tab:allsky} shows the number of detected short-period,
Jupiter-sized planets detected at $\sn\ge 20$ for four surveys: the SDSS-II supernova survey
\citep{sako06}, the Pan-STARRS\footnote{See http://pan-starrs.ifa.hawaii.edu/} 
medium-deep survey,
a Pan-STARRS survey with the same specifications as the medium-deep
survey, but covering 10 times the area (with the same amount of
time), and LSST assuming a 10-year survey covering 20,000 deg$^2$. 
Predictions are shown for both sun-like stars and M dwarfs (which can
be detected to much fainter magnitudes because the transit depths are
larger).  The potential of SDSS-II and the nominal Pan-STARRS survey
are small, due primarily to the fact they are targeted toward the
Galactic poles and are very deep, and so `run up' against the finite Galactic
scale height.  On the other hand, a wider Pan-STARRS survey spending
the same amount of time would detect at least twice as many planets.

The greatest potential comes from LSST, which could detect as many as
$\sim 8000$ transiting short-period planets around Sun-like stars, and
$\sim 15,000$ transiting planets around M-dwarfs, assuming the
frequency of short-period planets around M-dwarfs is the same as for
Sun-like stars.  Of course, culling all of these planets poses enormous
challenges.  Simply identifying the transiting planets themselves will
be difficult, due to the large number of trial periods that must be
searched given the $10~{\rm year}$ duration of observations.  More
worrisome, however, is the fact that these detections will likely
be associated with a much larger number of astrophysical false positives.  The large
number of expected candidates, combined with the fact that the
majority of the candidates will be quite faint ($V\ga 16$), implies
that it will be difficult to follow up every candidate individually to
exclude false positives.

Clearly much work needs to be done to realize the potential of large
scale synoptic surveys for the detection of transiting planets.  
The first step is to provide detailed simulations of the yield
of these large surveys.  These
simulations must include predictions for false positives, in order
to determine their expected contribution, as well as identify
observable trends that distinguish these
astrophysical backgrounds from the signal due to the planet population
of interest.  Then, it needs to be determined
whether methods can be devised to use these trends to
reliably separate the false positives from the bona fide
transiting planets, and so enable one to construct a statistical
sample of short-period transiting planets.

\acknowledgements
I would like to thank Thomas Beatty, Chris Burke, David Charbonneau,
Andrew Gould, Gabriella Mallen-Ornelas, Francis
O'Donovan, Josh Pepper, Frederic Pont, Sara Seager, and Andrzej Udalski for our
collaborations and/or invaluable discussions.  
I would particularly
like to thank Thomas Beatty for allowing me to present the results of
our paper prior to submission.  
I would like to thank the LOC,
Cristian Afonso, David Weldrake, and Maria Janssen-Bennynck, for
making this workshop possible.

\end{document}